# Synergistic Effect of Multi-Walled Carbon Nanotubes and Ladder-Type Conjugated Polymers on the Performance of N-Type Organic Electrochemical Transistors

*Silan Zhang, Matteo Massetti, Tero-Petri Ruoko, Deyu Tu, Chi-Yuan Yang, Xianjie Liu, Ziang Wu, Yoonjoo Lee, Renee Kroon, Per O. Å. Persson, Han Young Woo, Magnus Berggren, Christian Müller, Mats Fahlman, and Simone Fabiano\**

Organic electrochemical transistors (OECTs) have the potential to revolutionize the field of organic bioelectronics. To date, most of the reported OECTs include p-type (semi-)conducting polymers as the channel material, while n-type OECTs are yet at an early stage of development, with the best performing electron-transporting materials still suffering from low transconductance, low electron mobility, and slow response time. Here, the high electrical conductivity of multi-walled carbon nanotubes (MWCNTs) and the large volumetric capacitance of the ladder-type $\pi$-conjugated redox polymer poly(benzimidazobenzophenanthroline) (BBL) are leveraged to develop n-type OECTs with record-high performance. It is demonstrated that the use of MWCNTs enhances the electron mobility by more than one order of magnitude, yielding fast transistor transient response (down to 15 ms) and high $\mu C^*$ (electron mobility × volumetric capacitance) of about 1 F cm$^{-1}$ V$^{-1}$ s$^{-1}$. This enables the development of complementary inverters with a voltage gain of >16 and a large worst-case noise margin at a supply voltage of <0.6 V, while consuming less than 1 µW of power.

## 1. Introduction

Organic electrochemical transistors (OECTs) have undergone tremendous progress in recent years, primarily driven by their facile fabrication, biocompatibility, and low-voltage operation. To date, OECTs have proven their utility in a variety of electronic and bioelectronic applications, going from sensors[1] and medical diagnostics[2] to printable digital logic circuits[3] and non-volatile memories for neuromorphic computing,[4] to name just a few. This fast development has been enabled by a multidisciplinary research effort targeting new material design/synthesis,[5] device processing,[3c,6] and mixed ion-electron transport physics.[7]

In OECTs, an organic mixed ion-electron conductor forms the channel by bridging the source and drain electrodes and is in direct contact with an electrolyte. The application of a gate voltage ($V_G$) drives ions in the electrolyte inside the conducting polymer layer, leading to large changes in the source-drain current ($I_D$). Unlike field-effect transistors, the accumulation/depletion of charges will then occur throughout the entire bulk of the conducting polymer layer, with a consequently large volumetric capacitance ($C^*$). This makes OECTs effective ion-to-electron transducers.[8] The transconductance ($g_m = \partial I_D/\partial V_G$) is the figure of merit that quantifies the efficiency of this transduction, and depends on both the channel geometry and biasing conditions:[9]

S. Zhang, M. Massetti, T.-P. Ruoko, D. Tu, C.-Y. Yang, X. Liu,
R. Kroon, M. Berggren, M. Fahlman, S. Fabiano
Laboratory of Organic Electronics
Department of Science and Technology
Linköping University
Norrköping SE-601 74, Sweden
E-mail: simone.fabiano@liu.se
S. Zhang, R. Kroon, M. Berggren, M. Fahlman, S. Fabiano
Wallenberg Wood Science Center
Linköping University
Norrköping, Sweden

Z. Wu, Y. Lee, H. Y. Woo
Department of Chemistry
College of Science
Korea University
Seoul 136-713, Republic of Korea
P. O. Å. Persson
Thin Film Physics Division
Department of Physics
Chemistry and Biology (IFM)
Linköping University
Linköping SE-581 83, Sweden
C. Müller
Department of Chemistry and Chemical Engineering
and Wallenberg Wood Science Center
Chalmers University of Technology
Göteborg SE-412 96, Sweden

$$g_m = \frac{Wd}{L}\mu C^*(V_G - V_{th}) \quad (1)$$

where $L$ is the channel length, $W$ is the channel width, $d$ is the film thickness, $\mu$ is the electron mobility, and $V_{th}$ is the threshold voltage. Because of the volumetric charging/discharging effect, a geometry-normalized transconductance $g_{m,norm} = g_m \frac{L}{Wd}$ is often used when comparing the performance of different OECT channel materials.[5d,9]

While $g_{m,norm} > 0.6$ S cm$^{-1}$ have been achieved for several hole-transporting (p-type) organic conductors and semiconductors,[5c,7a,10] even the highest performing n-type (electron-transporting) OECT polymers lag far behind these values.[3b] This performance gap represents a major issue in the development of complementary circuits, since a balanced performance of both p- and n-type OECTs is needed. In addition, the n-type organic semiconductors should work stably in ambient and aqueous electrolytes, which requires that the lowest unoccupied molecular orbital energy level is deeper than −4.0 eV, a fundamental material design challenge.[11] Therefore, only a few n-type organic semiconductors have been reported to work effectively as n-channel materials in OECTs, to date. Thus, developing new n-type materials with improved OECT performances is of utmost importance.

Small-molecule fullerenes functionalized with hydrophilic glycolated side chains (C60-TEG) can achieve high electron mobility but suffer from low transconductance, high threshold voltage, and limited stability.[12] Naphthalenediimide (NDI)-based polymers bearing polar glycol side chains, like the donor-acceptor NDI-alkoxybithiophene (gT2) co-polymer P(gNDI-gT2),[13] have shown stable n-type OECT operation in water. However, NDI-based OECT materials like P(gNDI-gT2) typically suffer from a low electron mobility (<10$^{-5}$ cm$^2$ V$^{-1}$ s$^{-1}$), likely due to the highly localized nature of charge carriers and limited tolerance of the flexible polymer backbone towards disorder induced by intercalation of ions, which is typical for this class of polymers.[14] The low electron mobility would then limit the OECT transconductance and operational stability. Recently, it was shown that molecular n-doping can be used to produce OECTs with high transconductance and stable operation in water.[15] In particular, it was found that n-doping aids the electron mobility of the NDI-based polymer host, while favoring ion uptake and migration. However, despite the greatly improved mobility (≈2 × 10$^{-4}$ cm$^2$ V$^{-1}$ s$^{-1}$), n-doped NDI-based OECT polymers still suffer from low volumetric capacitance and thus, transconductance. Recently, rigid conjugated polymers bearing oligo(ethylene glycol) side chains were reported to achieve electron mobility of ≈6.5 × 10$^{-3}$ cm$^2$ V$^{-1}$ s$^{-1}$ and $g_{m,norm}$ of 0.21 S cm$^{-1}$, but suffered from slow response (≈127 ms) and limited operation stability.[16] The use of planar and rigid ladder-type polymers such as poly(benzimidazobenzophenanthroline) (BBL, **Figure 1**a) helps overcome this limitation, reaching record high transconductance and stable operation in water.[3b] The lack of solubilizing side chains results in large volumetric capacitance, exceeding that of NDI-based OECT polymers by more than twofold, but negatively impacts the device response time. While shortening the channel length down to 10 μm by photolithography enables BBL-based OECTs with fast temporal response of 5–10 ms,[17] the same polymer typically suffers from response times larger than 150 ms when implemented in printed OECTs with $L > 150$ μm.[18]

Here, we report on the synergistic effect of multi-walled carbon nanotubes (MWCNTs) and BBL on the performance of n-type OECTs. We demonstrate that the use of MWCNTs enhances the transistor transient response by over one order of magnitude, going from ≈200 ms of pristine BBL-based OECTs to ≈15 ms for BBL:MWCNT-based OECTs. We attribute the faster response time to the high conductivity of MWCNTs, which ensure a superior percolation path for electrons within the BBL channel, and thus high electron mobility and transconductance. The alcohol-based BBL:MWCNT dispersion can be deposited by simple spray-coating, a method which allows for large-area printing. We also developed electrochemical complementary inverters by combining BBL:MWCNT-based n-type OECTs with p-type OECTs comprising the hole-transporting bithiophene-thienothiophene copolymer P(g$_4$2T-TT) with tetraethylene glycol side chains. The resulting complementary OECT-based inverters show high voltage gains and large worst-case noise margin at a supply voltage <0.6 V, while consuming less than 1 μW of power.

## 2. Results and Discussion

Figure 1a,b shows the chemical structures of BBL and acid-treated MWCNTs. The MWCNTs were treated with a mixture of H$_2$SO$_4$ and HNO$_3$ (volumetric ratio 3:1) to aid their dispersion in solution, following a similar procedure reported earlier.[19] The acid treatment results in well-dispersed MWCNTs with an average diameter of about 13 nm and length of about ≈1–2 μm (**Figure 2**e and Figure S1, Supporting Information), and with a surface functionalized with carboxylic acid groups as shown by infrared spectroscopy measurements (see Figure S6, Supporting Information). BBL was dissolved in methanesulfonic acid (MSA), whereas the acid-treated MWCNTs were dispersed in isopropanol (IPA). Then, the BBL/MSA solution was slowly added to the MWCNT/IPA dispersion under rapid stirring, as shown in Figure 1c. As BBL is insoluble in IPA, it precipitates and wraps around the MWCNTs. The precipitate was washed with deionized water and IPA several times, and finally re-dispersed in IPA, resulting in a BBL:MWCNT dispersion with different mass ratios (100:1, 10:1, and 1:1). We refer to this BBL:MWCNT dispersion as "co-ink" to differentiate it from the BBL+MWCNT mixture (herein referred to as "mix-ink") in which a preformed BBL nanoparticle dispersion in IPA (i.e., BBL/IPA) was added to the MWCNT/IPA dispersion (Figure 1d). In analogy to the BBL:MWCNT co-ink, the mass ratio between BBL and MWCNT in the BBL+MWCNT mix-ink was varied from 100:1 to 1:1.

Then, we studied the film morphology of pristine BBL, acid-treated MWCNTs, BBL:MWCNT co-ink, and BBL+MWCNT mix-ink at different mass ratios by means of scanning electron microscopy (SEM). The morphology of spray-coated pristine BBL films is characterized by the presence of flakes forming a dense polymeric layer (Figure 2a), in agreement with previous reports.[3b] On the other hand, the presence of MWCNTs in the mixture results in a porous fibrillated thin-film morphology (Figure 2b–d), with BBL coating the MWCNT surface as suggested by the increase in fiber diameter with respect to the pristine MWCNTs (Figure 2e and Figure S2, Supporting Information). This is also corroborated by atomic force microscopy (AFM) (Figure S3, Supporting Information) and scanning transmission electron microscopy (STEM) measurements (Figure S4, Supporting Information), which show randomly dispersed MWCNTs inside the polymeric layers. In contrast, the SEM images of BBL+MWCNT mix-ink reported in Figure 2f–h show that MWCNTs are absorbed on the surface of BBL aggregates. In addition, when the mass ratio increases to 1:1 in the BBL+MWCNT mix-ink, a phase separation between BBL and MWCNT is visible in Figure 2h. SEM analysis of the buried interfaces (Figure S5, Supporting Information) shows that the bottom surface morphology of BBL:MWCNT mix-ink films matches that of the top surface (Figure 2f–h).

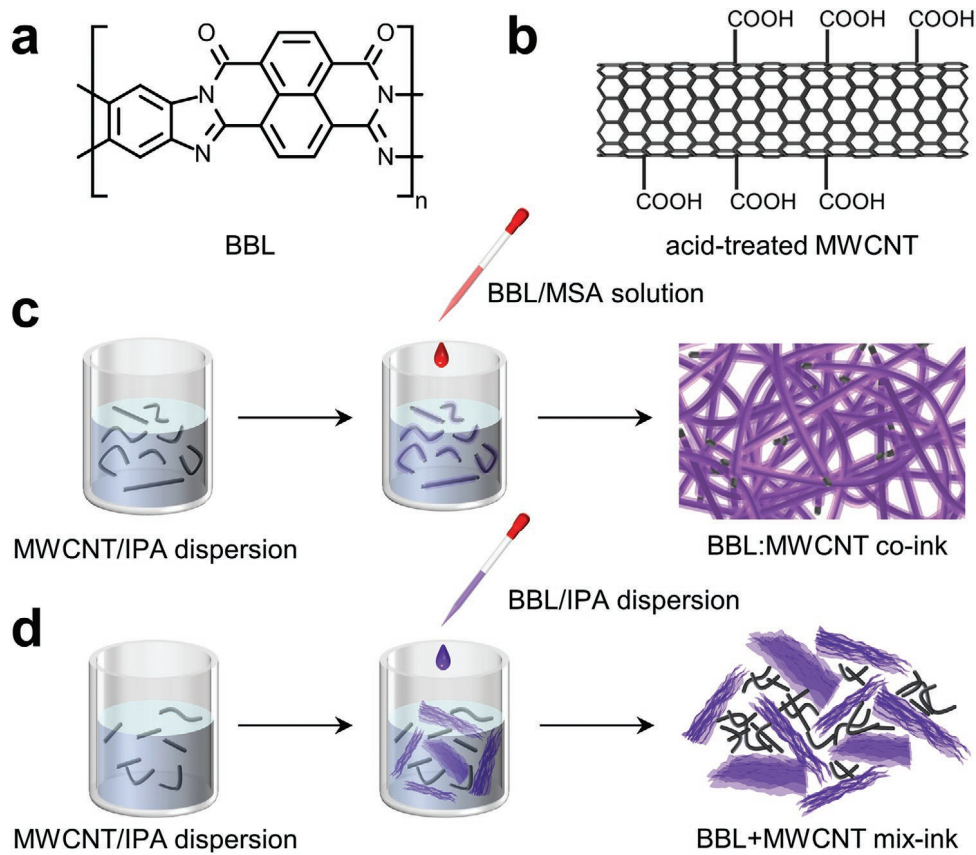

**Figure 1.** Chemical structures of a) BBL and b) acid-functionalized MWCNTs. c) Schematic illustration of the BBL:MWCNT co-ink preparation. d) Schematic illustration of the BBL+MWCNT mix-ink preparation.

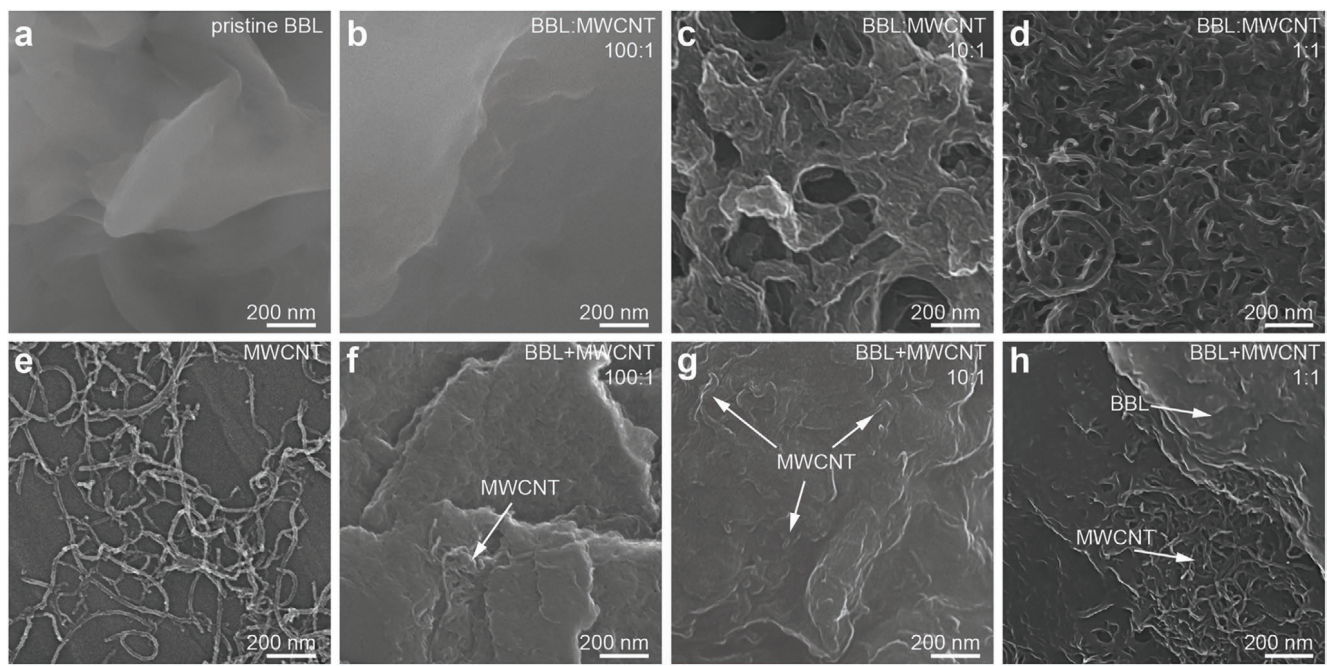

**Figure 2.** SEM images of a) pristine BBL and BBL:MWCNT co-ink at b) 100:1, c) 10:1, and d) 1:1 mass ratio. e) Acid treated MWCNTs and BBL+MWCNT mix-ink at f) 100:1, g) 10:1, and h) 1:1 mass ratio.

In order to rule out any chemical reaction between BBL and MWCNTs, we measured Fourier-transform infrared (FTIR) spectra of acid-treated MWCNTs, pristine BBL, and BBL:MWCNT composites (Figure S6, Supporting Information). Acid-treated MWCNTs show three absorption peaks at 1146 cm$^{-1}$ (C—O), 1569 cm$^{-1}$ (C=C), and 1716 cm$^{-1}$ (C=O), in agreement with previously reported spectra (Figure S6a, Supporting Information).[20] In the pristine BBL spectra, the peaks located at 1585 and 1703 cm$^{-1}$ are instead associated with the anti-symmetric C=C vibration and the C=O stretch, respectively (Figure S6b, Supporting Information). By increasing the amount of MWCNTs in the BBL:MWCNT blend, the peaks at 1174 cm$^{-1}$ (C—O in CNT overlapping with BBL peaks) and 1585 cm$^{-1}$ (C=C vibration peaks for both MWCNT and BBL) broaden and increase in amplitude without clear peak shifts, indicating that the composite spectra are formed as a superposition of the BBL and CNT contributions. This supports the assumption that BBL and CNT do not bond chemically with each other.

Grazing incidence wide-angle X-ray scattering (GIWAXS) was used to further compare the microstructures of BBL:MWCNT co-ink and BBL+MWCNT mix-ink composites (**Figure 3**). Pristine BBL chains are mainly oriented edge-on with respect to the substrate, with a lamellar (100) diffraction peak located at $q_z = 0.75$ Å$^{-1}$ (d-spacing = 8.37 Å) and a π–π (010) stacking peak located at $q_{xy} = 1.85$ Å$^{-1}$ (d-spacing = 3.39 Å). This molecular packing motif is in agreement with previous results.[21] MWCNT shows a strong peak at $q_z = 1.83$ Å$^{-1}$ relative to the (002) (labeled as (002)' to distinguish it from BBL's diffraction peaks) plane from graphite layers with a characteristic spacing of 3.43 Å.[22] BBL in both BBL:MWCNT co-ink and BBL+MWCNT mix-ink show the same edge-on orientation as in the pristine films. For low MWCNT contents (100:1 and 10:1), the diffraction of both co-ink and mix-ink is mainly dominated by BBL. When the mass ratio of BBL and MWCNT increases to 1:1, the diffraction pattern of BBL:MWCNT co-ink is mainly dominated by MWCNT, whereas the contribution of both BBL and MWCNT is visible in the diffraction pattern of BBL+MWCNT 1:1 mix-ink. This observation is in good agreement with the morphology observed in SEM. Indeed, in the case of BBL:MWCNT co-ink, BBL wraps around the MWCNTs, while in the case of BBL+MWCNT mix-ink, BBL and MWCNT aggregates are simply intermixed. To decouple the contribution of BBL and MWCNTs to the diffraction pattern, we fitted the peak located at $q_{xy} \approx 1.83$ Å$^{-1}$ for both co-ink and mix-ink films (see Figure S7, Supporting Information). The BBL/MWCNT peak area ratio decreases with increasing the content of MWCNTs for both co-ink and mix-ink films (Figure S7d, Supporting Information). We also estimated the average crystallite size based on Scherrer's equation using the full width at half maximum (FWHM) of the (100) peaks (Figure S8, Supporting Information). Both BBL:MWCNT co-ink and BBL+MWCNT mix-ink have a lower crystallinity compared to pristine BBL. However, at equal MWCNT content, BBL:MWCNT co-inks show higher crystallinity than BBL+MWCNT mix-ink.

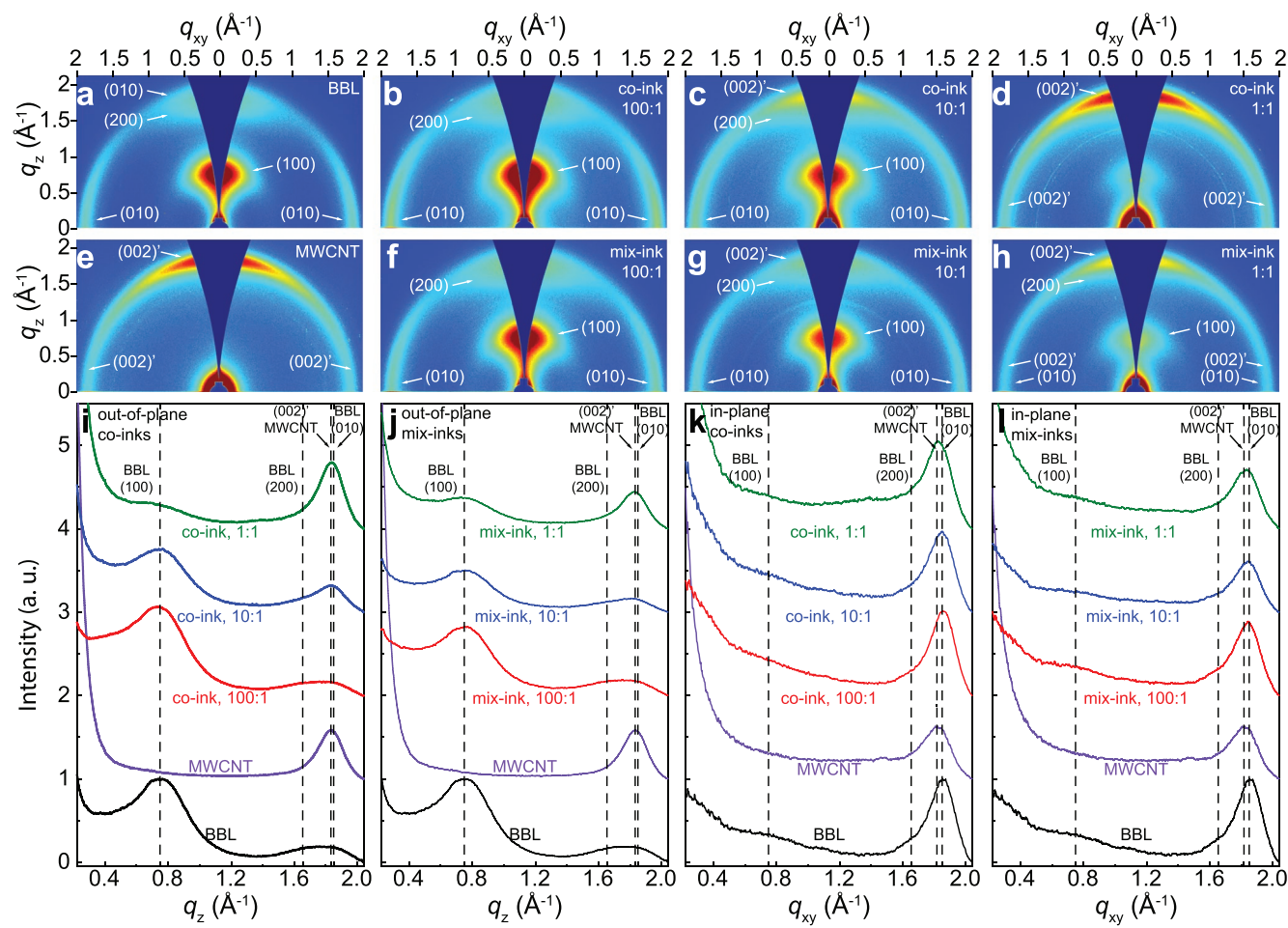

**Figure 3.** a–h) 2D GIWAXS patterns of pristine BBL (a), BBL:MWCNT co-ink (b–d), acid treated MWCNT (e), and BBL+MWCNT mix-ink (f–h). Corresponding 1D line cuts (i–l) in the i,j) out-of-plane and k,l) in-plane direction.

Ultraviolet photoelectron spectroscopy (UPS) was used to evaluate the work function (WF) changes of pristine BBL, MWCNTs, and BBL-MWCNT composites (Figure S9, Supporting Information). The UPS-derived WF is obtained from the low kinetic energy secondary electron cut-off onset, which yields a surface-area-averaged WF for the films studied.[23] The WF of pristine BBL is about 4.63 eV. The acid treatment significantly increases the WF of MWCNTs from 4.8 to 5.2 eV, most likely due to the presence of the surface dipoles deriving from the carboxylic functional groups.[24] For BBL:MWCNT co-ink, the WF gradually increases with increasing the amount of MWCNTs from 4.63 eV (pristine BBL) to 4.85 eV (BBL:MWCNT 1:1), which is indicative of well intermixed BBL and MWCNTs. In contrast, the WF of BBL+MWCNT 100:1 and 10:1 mix-ink is comparable to that of pristine BBL (4.5–4.6 eV) and sharply increases to about 5.1 eV for BBL+MWCNT 1:1. The mix-ink WF evolution suggests phase separation, with a BBL-dominated surface for the high BBL concentrations and an MWCNT-dominated surface for the 1:1 mix-ink formulation.

The electrical characteristics of BBL:MWCNT-based OECTs were evaluated in aqueous 0.1 M NaCl solution with a Ag/AgCl pellet as the gate electrode (**Figure 4**a). Figure 4b,c shows the transfer characteristics and corresponding $g_m$ for 40-nm-thick spray-coated films of BBL:MWCNT co-ink and BBL+MWCNT mix-ink, respectively. When a positive voltage is applied to the gate ($V_G$), cations penetrate the active layer while electrons are injected from the source, resulting in an increased source-drain current ($I_D$). The BBL:MWCNT-based OECTs transfer and output curves show typical accumulation-mode n-type transistor characteristics (Figure S10, Supporting Information). BBL:MWCNT 10:1 exhibits the highest performance with an increase in transconductance of 225% compared to pristine BBL OECTs and a maximum $g_m$ of about 45 µS (Figure 4b). Note that the gate current ($I_G$) is more than one order of magnitude smaller than $I_D$ at $V_G$ >0.4 V (Figure S11, Supporting Information). At higher MWCNT load, BBL:MWCNT 1:10 and pure MWCNTs show no OECT performance, with the drain current purely dominated by the leakage current (Figure S12, Supporting Information). In contrast, the OECT transfer characteristics of BBL+MWCNT mix-ink show a continuous decrease of $I_D$ and $g_m$ with increasing the amount of MWCNT in the mixture (Figure 4c). We then argue that the increase in $g_m$ observed for BBL:MWCNT co-ink is not due to a shortening of the effective channel, but instead to a superior percolation path for electrons within the BBL-wrapped MWCNTs (see mobility discussion below).

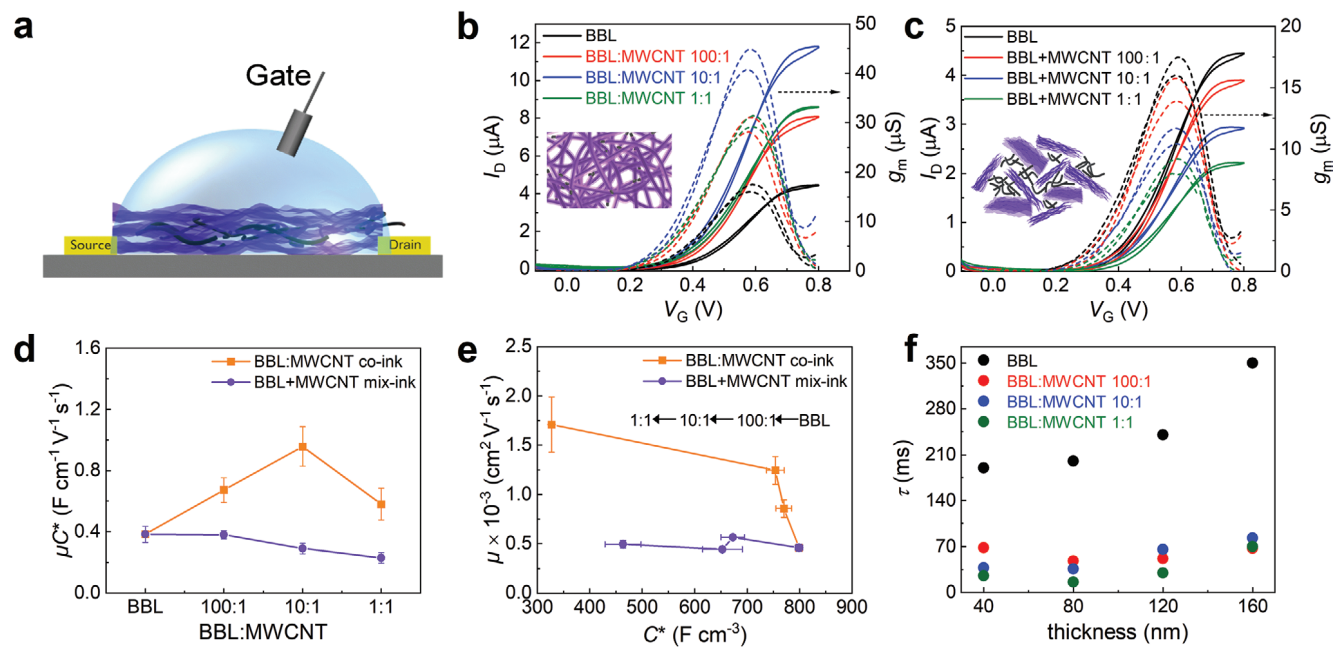

**Figure 4.** OECTs performances with different BBL:MWCNT composites. a) Schematic illustration of OECTs device structure. Transfer curves (solid line) and corresponding transconductance ($g_m$, dash line) at $V_D$ = 0.6 V ($L$ = 30 µm, $W$ = 1000 µm, $d$ = 40 nm) of b) BBL:MWCNT co-ink and c) BBL+MWCNT mix-ink. d) Comparison of $\mu C^*$ with BBL-MWCNT co-ink and mix-ink. e) Electron mobility versus volumetric capacitance in BBL-MWCNT co-ink and mix-ink. f) Switching speed for different BBL:MWCNT co-ink film thickness.

The thickness of the spray-coated BBL:MWCNT co-ink films was varied from 40 to 160 nm by controlling the concentration of the sprayed ink (Figure S13, Supporting Information). Because charge accumulation occurs throughout the entire OECT channel material, the transconductance increases linearly with the semiconductor film thickness (Figure S14, Supporting Information). In order to compare the performance of our BBL:MWCNT-based OECTs with other benchmark n-type OECT materials, we normalized $g_m$ by the geometry factor $\frac{Wd}{L}$. BBL:MWCNT 10:1 shows the highest $g_{m,\,norm} = (0.30 \pm 0.03)$ S cm$^{-1}$ (max $g_{m,\,norm} = 0.34$ S cm$^{-1}$), which is also among the highest values reported for n-type OECTs (see Table S1, Supporting Information).

As the OECT transconductance is proportional to the product of $\mu C^*$, we investigated how the amount of MWCNTs influences $\mu$ and $C^*$, separately. Figure 4d shows that the benchmark figure-of-merit $\mu C^*$ first increases from $0.38 \pm 0.05$ F cm$^{-1}$ V$^{-1}$ s$^{-1}$ for pure BBL to $0.96 \pm 0.13$ F cm$^{-1}$ V$^{-1}$ s$^{-1}$ for BBL:MWCNT 10:1 co-ink, and then decreases to $0.58 \pm 0.10$ F cm$^{-1}$ V$^{-1}$ s$^{-1}$ for the 1:1 mass ratio. Opposite trend is observed for the $\mu C^*$ of BBL+MWCNT mix-ink, which monotonously decreases with increasing the amount of MWCNT in the mixture, reaching $0.23 \pm 0.04$ F cm$^{-1}$ V$^{-1}$ s$^{-1}$ for BBL+MWCNT 1:1 mix-ink.

We then used electrochemical impedance spectroscopy (EIS) to quantify the volumetric capacitance $C^*$. Pristine BBL, being a redox conjugated polymer, shows a pair of redox peaks located at –0.7 and –0.8 V versus Ag/AgCl (Figure S15, Supporting Information). On the contrary, metallic MWCNTs are not redox-active and show a very low electric double-layer capacitance. Hence, BBL dominates $C^*$ of BBL:MWCNT films, and any increase in MWCNT load results in a decrease of the capacitance regardless of the type of ink. Pristine BBL has a volumetric capacitance of $799 \pm 7$ F cm$^{-3}$, which only decreases by 4–5% for BBL:MWCNT 100:1 ($770 \pm 15$ F cm$^{-3}$) and BBL:MWCNT 10:1 ($754 \pm 17$ F cm$^{-3}$) co-ink (measured at –0.7 V versus Ag/AgCl, Figure 4e and Figures S16 and S17, Supporting Information). While the $C^*$ of BBL:MWCNT decreases slightly with increasing the CNT content, it remains 2–7 times higher than that of NDI-based polymers and fullerene-derivatives (see Table S1, Supporting Information). However, when the BBL:MWCNT mass ratio reaches 1:1, $C^*$ drops to $327 \pm 5$ F cm$^{-3}$, which would explain the lower $g_m$ value of BBL:MWCNT 1:1 as compared to BBL:MWCNT 10:1. In the case of BBL+MWCNT mix-ink, we observed a more pronounced decrease in $C^*$, with values dropping by more than 18% when going from pristine BBL to BBL+MWCNT 10:1 mix-ink ($653 \pm 38$ F cm$^{-3}$). We attributed the slightly larger $C^*$ of BBL:MWCNT co-ink to the larger surface area, which aids the diffusion of ions inside the film bulk and counterbalance the reduction of redox active polymer in the blend.

Having measured $\mu C^*$ and $C^*$ independently, we studied the evolution of electron mobility with MWCNT content (see Figure 4e). The electron mobility of BBL:MWCNT co-ink increases almost linearly with the MWCNT content, while an opposite trend is observed for BBL+MWCNT mix-ink. For pristine BBL, the extracted electron mobility is $(0.5 \pm 0.03) \times 10^{-3}$ cm$^2$ V$^{-1}$ s$^{-1}$ and increases about 400%, reaching a value of $(1.7 \pm 0.3) \times 10^{-3}$ cm$^2$ V$^{-1}$ s$^{-1}$, for BBL:MWCNT 1:1. Although the $\mu$ of BBL:MWCNT 1:1 is up to 4 times higher than that of pristine BBL, its $C^*$ is less than half that of the pure polymer. Therefore, the $g_m$ of BBL:MWCNT 1:1 is only 1.5 times higher than pristine BBL. BBL:MWCNT 10:1, which has the highest $g_m$, shows an electron mobility of $(1.2 \pm 0.14) \times 10^{-3}$ cm$^2$ V$^{-1}$ s$^{-1}$. Note that we also estimated the charge carrier mobility through the constant gate current[7c] and impedance matching[25] methods and observed similar trends (Figure S18, Supporting Information). These mobility values are 1–3 orders of magnitude higher than those reported for NDI-based polymers used in OECTs (Table S1, Supporting Information), and on par with thiophene-based polymers used in accumulation-mode transistors ($\approx 1.3 \times 10^{-3}$ cm$^2$ V$^{-1}$ s$^{-1}$).[5c,13,15,26] In the case of BBL+MWCNT mix-ink, the electron mobility is observed to be almost independent of the MWCNT content (average mobility $\approx 0.5 \times 10^{-3}$ cm$^2$ V$^{-1}$ s$^{-1}$). We interpreted the increase in electron mobility observed for BBL:MWCNT co-ink in terms of a superior percolation path for electrons provided by the MWCNTs. For this reason, we only focus on BBL:MWCNT co-ink in the rest of the study.

To understand the influence of MWCNTs on the electrode/electrolyte interface kinetics, we analyzed the charge transfer resistance ($R_{ct}$) from the Nyquist plots. As reported in Figure S19a, Supporting Information, it can be seen that the $R_{ct}$ decreases with increasing amount of MWCNTs in the BBL:MWCNT co-ink, going from 47 Ω for pristine BBL to 30 Ω for BBL:MWCNT (1:1). This is consistent with the increased conductivity due to the presence of MWCNTs, and also indicative of a faster ion transfer at the BBL/electrolyte interface due to the large film porosity. In contrast, the $R_{ct}$ extracted for BBL+MWCNT mix-ink increases to 58 Ω for BBL+MWCNT (1:1), which is indicative of a slower ion transfer at the polymer/electrolyte interface (Figure S19b). These results also suggest that BBL and MWCNT are better intermixed in the BBL:MWCNT co-ink compared to BBL+MWCNT mix-ink, thus to facilitate the charge (ion/electron) transfer process.

In order to investigate the effect of MWCNTs on the otherwise slow transient response of BBL-based OECTs,[3b,11b,15] we measured the switching speed of BBL:MWCNT-based OECTs at different MWCNT contents. The switching time ($\tau$) is estimated by pulsing the gate voltage to 0.6 V (pulse length 2.5 s) at a constant drain voltage ($V_D = 0.6$ V) and fitting the current to an exponential function (see Figure S20, Supporting Information). As shown in Figure 4f, the switching speed strongly varies with the channel thickness. For pristine BBL, $\tau$ is equal to $(190 \pm 16)$ ms for a 40-nm-thick film and increases to $(350 \pm 29)$ ms when the film thickness is increased to 160 nm. This is indicative of a limited ion diffusion into the OECT channel.[3b] Increasing the amount of MWCNTs gradually decreases $\tau$ (Figure 4f and Figure S20, Supporting Information), as a consequence of the increase in mobility. In addition to this, the coarse BBL:MWCNT structure observed in the SEM images displayed in Figure 2b–d helps the electrolyte diffuse inside the channel, also contributing to the faster switching speed. The BBL:MWCNT (1:1) shows the fastest transient response with a $\tau$ as low as (16 ± 1) ms for an 80-nm-thick film, which is one order of magnitude faster than a pristine BBL film of the same thickness and 2–5 times faster than n-doped NDI-based polymers[15] and small-molecule glycolated fullerenes[12] (see Table S1, Supporting Information). In addition, no current degradation is observed while continuously operating the devices in water for 2 h (Figure S21, Supporting Information). Note also that the co-ink dispersions are very stable when stored in ambient, and show no degradation of the electrical performance even after 1 year (Figure S22, Supporting Information). We also fabricated screen-printed OECTs with channel dimensions of $L = 200$ μm, $W = 2000$ μm, and channel thickness $d = 40$ nm, which are larger than typical photolithography-made or thermally-evaporated source/drain electrode OECTs.[3a] The screen-printed large channel OECTs exhibited a drop in $\tau$ from 140 ms (pristine BBL) to 35 ms (BBL:MWCNT 1:1) (Figure S23, Supporting Information), which paves the way for the development of faster complementary screen-printed circuits.

We fabricated complementary inverters by combining the best performing BBL:MWCNT-based n-type OECTs with p-type OECTs based on P(g$_4$2T-TT). Complementary inverters have the advantage of low static power consumption and large noise margins.[27] The gain of the inverter, extracted from the first derivate of the voltage-transfer characteristics (gain = $\partial V_{out}/\partial V_{in}$), is proportional to transistor transconductance and output resistance. Thus, in order to achieve higher gain, one can either increase $g_m$ or the output resistance by changing the channel geometry. However, this strategy also increases the parasitic capacitance which limits the inverter bandwidth.[28] An alternative way to obtain a high gain is to increase the semiconductor mobility. Therefore, we tested the impact of MWCNTs on the inverter characteristics (**Figure 5**). For the p-type OECT, we used P(g$_4$2T-TT), a benchmark p-type polymer for accumulation-mode OECTs (see inset to Figure 5a for its chemical structure). Figure 5a shows the typical transfer characteristics of a 40-nm-thick P(g$_4$2T-TT)-based OECT operated in 0.1 M NaCl with $L = 30$ μm and $W = 1000$ μm. The P(g$_4$2T-TT)-based OECT is then connected by an external wire to a 160 nm-thick BBL-based OECT. The voltage-transfer characteristics of the inverters at different MWCNT content are shown in Figure S24, Supporting Information. The gain increases with the amount of MWCNTs and reaches a value of about 16 for BBL:MWCNT (10:1) for $V_{DD} = 0.6$ V (see Figure 5b). These gain values are about 2.3 times higher than those of pristine BBL and in good agreement with the observed increase in $g_m$. The evolution of the gain values with $V_{DD}$, at different MWCNT content, is shown in Figure 5c. In addition, all the inverters show a dynamic power consumption lower than 1 μW, and BBL:MWCNT (10:1) also shows the lowest static-state power consumption of about 0.04 μW (Figure S25, Supporting Information).

Complementary inverters allow for a high noise margin, that is, the level of noise an inverter can tolerate without compromising its logic operation. As shown in Figure S26, Supporting Information, the lowest/highest input voltage ($V_{IL}$, $V_{IH}$) and the corresponding highest/lowest output voltage ($V_{OH}$, $V_{OL}$) can be extracted from the voltage-transfer characteristic curves at gain = $\partial V_{out}/\partial V_{in} = -1$. Figure S27, Supporting Information, and Figure 5d shows the noise margin high (NM$_H$) and noise margin low ($NM_L$) and corresponding percentages of maximum theoretical value $V_M = V_{DD}/2$. Since $NM$ is proportional to $-1/$(gain), BBL:MWCNT 10:1 based inverters show the highest gain as well as the highest $NM_L$ and $NM_H$.

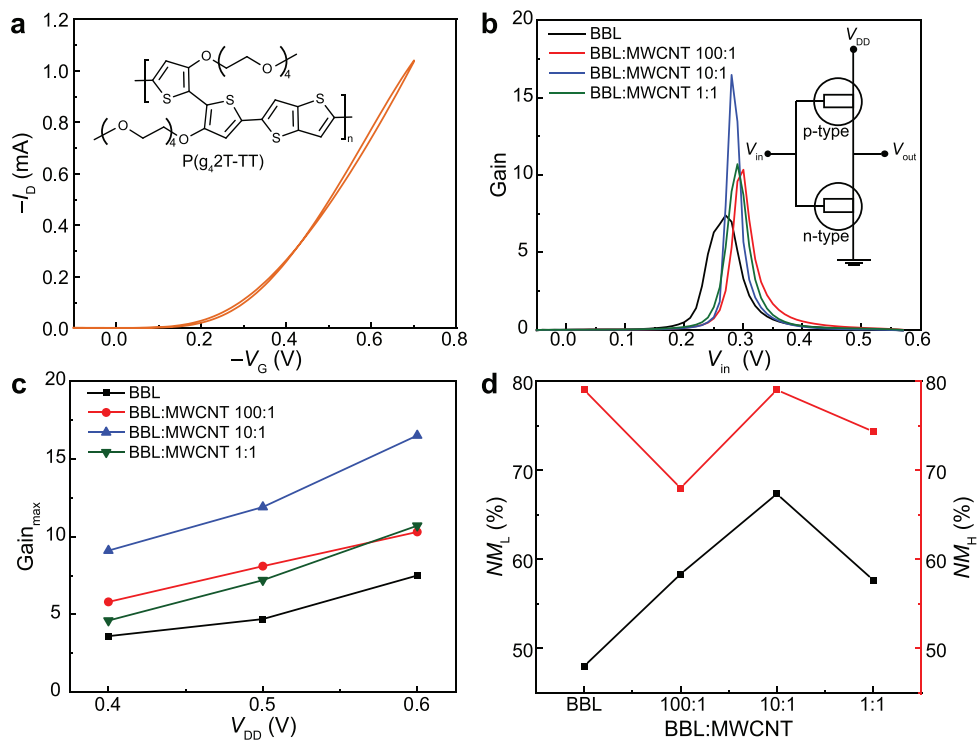

**Figure 5.** The complementary inverters characteristics. a) Transfer and P(g$_4$2T-TT) chemical structure (inset) of p-type (P(g$_4$2T-TT)) OECT ($L = 30$ μm, $W = 1000$ μm, $d = 40$ nm). b) Voltage gains with different BBL:MWCNT at $V_{DD} = 0.6$ V (inset: circuit diagram of an inverter). c) Comparison of the maximum gains with different BBL:MWCNT mass ratio at different $V_{DD}$. d) The noise margin low (NM$_L$) and noise margin high (NM$_H$) percentages of the maximum theoretical value ($V_{DD}/2 = 0.3$ V) of different BBL:MWCNT mass ratio devices.

## 3. Conclusions

In conclusion, we leveraged the high electrical conductivity of MWCNTs and the large volumetric capacitance of the ladder-type π-conjugated redox polymer BBL to develop n-type OECTs with record-high performance. We demonstrated that the use of MWCNTs enhances the transistor transient response by more than one order of magnitude, approaching ≈15 ms for BBL:MWCNT-based OECTs. We attributed the faster response time to the high electrical conductivity of MWCNTs, which yields an efficient percolation path for charges within the BBL channel. This enables high OECT electron mobility >10$^{-3}$ cm$^2$ V$^{-1}$ s$^{-1}$ which, combined with the large BBL volumetric capacitance (>700 F cm$^{-3}$), results in a maximum normalized transconductance of up to 0.34 S cm$^{-1}$. This is among the highest transconductance reported to date for n-type OECTs, and on the same order of magnitude as corresponding values reported for p-type thiophene polymer-based OECTs (≈0.60 S cm$^{-1}$).[10] We also fabricated complementary inverters by combining BBL:MWCNT-based n-type OECTs with p-type P(g$_4$2T-TT)-based OECTs reaching a voltage gain > 16 and large worst-case noise margin at a supply voltage <0.6 V, while consuming less than 1 µW of power. The alcohol-based BBL:MWCNT dispersion can be deposited by simple spray-coating, a method which is compatible with other large-area printing techniques, thus paving the way toward fully-printed complementary logics and sensors for the Internet of Things and bioelectronics applications.

## 4. Experimental Section

*Acid-Treated MWCNTs*: MWCNTs (>95%, outer diameter: 6–9 nm, length: 5 µm), sulfuric acid (H$_2$SO$_4$), and nitric acid (HNO$_3$) were purchased from Sigma-Aldrich. For acid treatment, 0.5 g MWCNTs were mixed with 150 mL concentrated H$_2$SO$_4$ and 50 mL HNO$_3$ stirred at 80 °C overnight. The acid-treated MWCNTs were then washed by DI water several times until the supernatant pH was around 7.

*Dispersion Preparation*: The dispersion preparation is illustrated in Figure 1 and in accordance to previous report.[29] In brief, the acid-treated MWCNTs were firstly dispersed in MSA (MSA, >99%, purchased from Sigma-Aldrich) to form a 2 mg mL$^{-1}$ dispersion. BBL (purchased from Sigma-Aldrich) were dissolved in MSA to form a 2 mg mL$^{-1}$ solution. Part of the MWCNTs dispersion was dispersed in IPA and then the BBL/MSA solution was added to the dispersion slowly under rapid stirring. The mixed dispersion was washed with deionized water and IPA by centrifugation until the supernatant was no longer acidic. The BBL:MWCNT was then dispersed in IPA. The pristine BBL dispersion was prepared with the same method without MWCNTs.

*Device Fabrication*: All devices were fabricated on glass substrates cleaned with sequential sonication in 2% Hellmanex, deionized water, acetone, and IPA and followed by drying with nitrogen. Source and drain electrodes (5 nm Cr and 50 nm Au) were deposited by physical vapor deposition using an evaporation mask. BBL-based materials were spray-coated on the top of glasses by commercial air brush. Since the concentration of dispersions may varied, 30–50 µL dispersion was diluted in 1 mL IPA for 40 nm thick film (thickness was measured by Dektak surface profilometer). Thickness was controlled by adding different aliquots of the BBL-based dispersion to IPA. After spray-coating, the films were immersed in deionized water to make sure that MSA was removed completely and then followed by drying the film on a hot plate at 150 °C for 10 min. For p-type OECTs, P(g$_4$2T-TT) was synthesized as reported previously.[30] For 40 nm thick film, 5 mg mL$^{-1}$ P(g$_4$2T-TT) was dissolved in chlorobenzene and spin-coated at 1000 rpm for 1 min then dried at 100 °C. For all the OECTs, an Ag/AgCl pellet electrode (Warner Instruments, USA) was used as the gate.

*Materials Characterization*: Morphology characterizations were performed by SEM (ZEISS Sigma 500) at 20 kV and AFM (Veeco Instruments Dimension 3100) in tapping mode. The FTIR spectra (Bruker Equinox 55) were measured in transmission mode. STEM was performed using the double corrected Linköping Titan electron microscope, as described elsewhere,[31] operated at 300 kV. 2D GIWAXS patterns were collected from 100 nm-thick films spray-coated on Si wafer substrate. Scattering was carried out at Beamline 9A at the Pohang Accelerator Laboratory in South Korea with 11.07 eV synchrotron radiation, with a 0.14° incident angle. The exposure time was 10 s.

*Electrical and Electrochemical Characterizations*: All the OECTs characteristics were tested by Keithley 4200 semiconductor parameter analyzer. For the estimation of the mobility through the constant gate current method:[7c] the drain current transients (d$I_D$/d$t$) were extracted for a constant $I_G$ and at a particular $V_D$. The electron transit time ($\tau_e$) was calculated from d$I_D$/d$t = -I_G/\tau_e$. The electron transit time $\tau_e$ was also extracted by using the impedance matching method.[25] Constant drain bias was applied to the channel and a sinusoidal voltage signal at a gate electrode with a preset offset and frequency. The frequency-domain relation between the measured gate current ($\Delta I_G$) and drain current ($\Delta I_D$) is: $\Delta I_G(f) = 2\pi f \tau_e \Delta I_D(f)$. Using the derived $\tau_e$, the mobility $\mu$ can be extracted from $\mu = \frac{L^2}{\tau_e V_D}$, where $L$ is channel length. The cyclic voltammetry and EIS were characterized by Bio-Logic Science Instruments. For all the electrochemical measurement, BBL:MWCNT electrode was spray-coated on 1 × 1 cm$^2$ Cr/Au electrode on glass substrate.

## Supporting Information

Supporting Information is available from the Wiley Online Library or from the author.

## Acknowledgements

The authors wish to thank Mikhail Vagin and Hengda Sun (Linköping University) for insightful discussions, as well as Qilun Zhang, Xiane Li, and Yonzhen Chen (Linköping University) for assistance with UPS measurement. This work was financially supported by the Knut and Alice Wallenberg Foundation, the Swedish Research Council (2016-03979, 2020-03243), Swedish Foundation for Strategic Research (14-0074), ÅForsk (18-313, 19-310), Olle Engkvists Stiftelse (204-0256), the Swedish Government Strategic Research Area in Materials Science on Functional Materials at Linköping University (Faculty Grant SFO-Mat- LiU 2009-00971), the European Commission through the FET-OPEN project MITICS (GA-964677), and VINNOVA (2020-05223). H.Y.W. acknowledges financial support from the National Research Foundation of Korea (NRF-2019R1A2C2085290, 2019R1A6A1A11044070).

## Conflict of Interest

The authors declare no conflict of interest.

## Data Availability Statement

The data that support the findings of this study are available from the corresponding author upon reasonable request.

[1] a) P. Lin, F. Yan, J. Yu, H. L. W. Chan, M. Yang, *Adv. Mater.* **2010**, *22*, 3655; b) A. M. Pappa, D. Ohayon, A. Giovannitti, I. P. Maria, A. Savva, I. Uguz, J. Rivnay, I. McCulloch, R. M. Owens, S. Inal, *Sci. Adv.* **2018**, *4*, eaat0911; c) S. Wustoni, C. Combe, D. Ohayon, M. H. Akhtar, I. McCulloch, S. Inal, *Adv. Funct. Mater.* **2019**, *29*, 1904403.

[2] a) S. Park, S. W. Heo, W. Lee, D. Inoue, Z. Jiang, K. Yu, H. Jinno, D. Hashizume, M. Sekino, T. Yokota, K. Fukuda, K. Tajima, T. Someya, *Nature* **2018**, *561*, 516; b) D. Khodagholy, T. Doublet, P. Quilichini, M. Gurfinkel, P. Leleux, A. Ghestem, E. Ismailova, T. Hervé, S. Sanaur, C. Bernard, G. G. Malliaras, *Nat. Commun.* **2013**, *4*, 1575; c) P. Leleux, J. Rivnay, T. Lonjaret, J.-M. Badier, C. Bénar, T. Hervé, P. Chauvel, G. G. Malliaras, *Adv. Healthcare Mater.* **2015**, *4*, 142; d) M. Braendlein, T. Lonjaret, P. Leleux, J.-M. Badier, G. G. Malliaras, *Adv. Sci.* **2017**, *4*, 1600247.

[3] a) P. Andersson Ersman, R. Lassnig, J. Strandberg, D. Tu, V. Keshmiri, R. Forchheimer, S. Fabiano, G. Gustafsson, M. Berggren, *Nat. Commun.* **2019**, *10*, 5053; b) H. Sun, M. Vagin, S. Wang, X. Crispin, R. Forchheimer, M. Berggren, S. Fabiano, *Adv. Mater.* **2018**, *30*, 1704916; c) M. Hamedi, R. Forchheimer, O. Inganäs, *Nat. Mater.* **2007**, *6*, 357.

[4] a) J. Y. Gerasimov, R. Gabrielsson, R. Forchheimer, E. Stavrinidou, D. T. Simon, M. Berggren, S. Fabiano, *Adv. Sci.* **2019**, *6*, 1801339; b) P. Gkoupidenis, D. A. Koutsouras, G. G. Malliaras, *Nat. Commun.* **2017**, *8*, 15448; c) P. Gkoupidenis, N. Schaefer, B. Garlan, G. G. Malliaras, *Adv. Mater.* **2015**, *27*, 7176; d) S. Yamamoto, G. G. Malliaras, *ACS Appl. Electron. Mater.* **2020**, *2*, 2224.

[5] a) M. Moser, A. Savva, K. Thorley, B. D. Paulsen, T. C. Hidalgo, D. Ohayon, H. Chen, A. Giovannitti, A. Marks, N. Gasparini, A. Wadsworth, J. Rivnay, S. Inal, I. McCulloch, *Angew. Chem., Int. Ed.* **2021**, *60*, 7777; b) R. H. Karlsson, A. Herland, M. Hamedi, J. A. Wigenius, A. Åslund, X. Liu, M. Fahlman, O. Inganäs, P. Konradsson, *Chem. Mater.* **2009**, *21*, 1815; c) S. Inal, J. Rivnay, P. Leleux, M. Ferro, M. Ramuz, J. C. Brendel, M. M. Schmidt, M. Thelakkat, G. G. Malliaras, *Adv. Mater.* **2014**, *26*, 7450; d) A. Giovannitti, R. B. Rashid, Q. Thiburce, B. D. Paulsen, C. Cendra, K. Thorley, D. Moia, J. T. Mefford, D. Hanifi, D. Weiyuan, M. Moser, A. Salleo, J. Nelson, I. McCulloch, J. Rivnay, *Adv. Mater.* **2020**, *32*, 1908047; e) E. Zeglio, O. Inganäs, *Adv. Mater.* **2018**, *30*, 1800941.

[6] a) M. Zabihipour, R. Lassnig, J. Strandberg, M. Berggren, S. Fabiano, I. Engquist, P. Andersson Ersman, *npj Flexible Electron.* **2020**, *4*, 15; b) B. Schmatz, A. W. Lang, J. R. Reynolds, *Adv. Funct. Mater.* **2019**, *29*, 1905266; c) C. Qian, J. Sun, L.-a. Kong, G. Gou, J. Yang, J. He, Y. Gao, Q. Wan, *ACS Appl. Mater. Interfaces* **2016**, *8*, 26169; d) B. Marchiori, R. Delattre, S. Hannah, S. Blayac, M. Ramuz, *Sci. Rep.* **2018**, *8*, 8477.

[7] a) J. T. Friedlein, R. R. McLeod, J. Rivnay, *Org. Electron.* **2018**, *63*, 398; b) D. Tu, S. Fabiano, *Appl. Phys. Lett.* **2020**, *117*, 080501; c) D. A. Bernards, G. G. Malliaras, *Adv. Funct. Mater.* **2007**, *17*, 3538; d) L. Q. Flagg, R. Giridharagopal, J. Guo, D. S. Ginger, *Chem. Mater.* **2018**, *30*, 5380; e) A. Savva, R. Hallani, C. Cendra, J. Surgailis, T. C. Hidalgo, S. Wustoni, R. Sheelamanthula, X. Chen, M. Kirkus, A. Giovannitti, A. Salleo, I. McCulloch, S. Inal, *Adv. Funct. Mater.* **2020**, *30*, 1907657; f) B. D. Paulsen, S. Fabiano, J. Rivnay, *Annu. Rev. Mater. Res.* **2021**, *51*, 73.

[8] D. Khodagholy, J. Rivnay, M. Sessolo, M. Gurfinkel, P. Leleux, L. H. Jimison, E. Stavrinidou, T. Herve, S. Sanaur, R. M. Owens, G. G. Malliaras, *Nat. Commun.* **2013**, *4*, 2133.

[9] S. Inal, G. G. Malliaras, J. Rivnay, *Nat. Commun.* **2017**, *8*, 1767.

[10] C. B. Nielsen, A. Giovannitti, D.-T. Sbircea, E. Bandiello, M. R. Niazi, D. A. Hanifi, M. Sessolo, A. Amassian, G. G. Malliaras, J. Rivnay, I. McCulloch, *J. Am. Chem. Soc.* **2016**, *138*, 10252.

[11] a) X. Zhan, A. Facchetti, S. Barlow, T. J. Marks, M. A. Ratner, M. R. Wasielewski, S. R. Marder, *Adv. Mater.* **2011**, *23*, 268; b) H. Sun, J. Gerasimov, M. Berggren, S. Fabiano, *J. Mater. Chem. C* **2018**, *6*, 11778; c) S. Griggs, A. Marks, H. Bristow, I. McCulloch, *J. Mater. Chem. C* **2021**, *9*, 8099.

[12] C. G. Bischak, L. Q. Flagg, K. Yan, C.-Z. Li, D. S. Ginger, *ACS Appl. Mater. Interfaces* **2019**, *11*, 28138.

[13] A. Giovannitti, C. B. Nielsen, D.-T. Sbircea, S. Inal, M. Donahue, M. R. Niazi, D. A. Hanifi, A. Amassian, G. G. Malliaras, J. Rivnay, I. McCulloch, *Nat. Commun.* **2016**, *7*, 13066.

[14] a) S. Wang, H. Sun, T. Erdmann, G. Wang, D. Fazzi, U. Lappan, Y. Puttisong, Z. Chen, M. Berggren, X. Crispin, A. Kiriy, B. Voit, T. J. Marks, S. Fabiano, A. Facchetti, *Adv. Mater.* **2018**, *30*, 1801898; b) S. Wang, D. Fazzi, Y. Puttisong, M. J. Jafari, Z. Chen, T. Ederth, J. W. Andreasen, W. M. Chen, A. Facchetti, S. Fabiano, *Chem. Mater.* **2019**, *31*, 3395.

[15] A. F. Paterson, A. Savva, S. Wustoni, L. Tsetseris, B. D. Paulsen, H. Faber, A. H. Emwas, X. Chen, G. Nikiforidis, T. C. Hidalgo, M. Moser, I. P. Maria, J. Rivnay, I. McCulloch, T. D. Anthopoulos, S. Inal, *Nat. Commun.* **2020**, *11*, 3004.

[16] X. Chen, A. Marks, B. D. Paulsen, R. Wu, R. B. Rashid, H. Chen, M. Alsufyani, J. Rivnay, I. McCulloch, *Angew. Chem., Int. Ed.* **2021**, *60*, 9368.

[17] J. Surgailis, A. Savva, V. Druet, B. D. Paulsen, R. Wu, A. Hamidi-Sakr, D. Ohayon, G. Nikiforidis, X. Chen, I. McCulloch, J. Rivnay, S. Inal, *Adv. Funct. Mater.* **2021**, *31*, 2010165.

[18] D. T. Chi-Yuan Yang, T.-P. Ruoko, J. Y. Gerasimov, H.-Y. Wu, P. C. Harikesh, R. Kroon, C. Müller, M. Berggren, S. Fabiano, arXiv:2106.07438 **2021**.

[19] J. Zhang, H. Zou, Q. Qing, Y. Yang, Q. Li, Z. Liu, X. Guo, Z. Du, *J. Phys. Chem. B* **2003**, *107*, 3712.

[20] F. Avilés, J. V. Cauich-Rodríguez, L. Moo-Tah, A. May-Pat, R. Vargas-Coronado, *Carbon* **2009**, *47*, 2970.

[21] a) A. L. Briseno, S. C. B. Mannsfeld, P. J. Shamberger, F. S. Ohuchi, Z. Bao, S. A. Jenekhe, Y. Xia, *Chem. Mater.* **2008**, *20*, 4712; b) C.-Y. Yang, M.-A. Stoeckel, T.-P. Ruoko, H.-Y. Wu, X. Liu, N. B. Kolhe, Z. Wu, Y. Puttisong, C. Musumeci, M. Massetti, H. Sun, K. Xu, D. Tu, W. M. Chen, H. Y. Woo, M. Fahlman, S. A. Jenekhe, M. Berggren, S. Fabiano, *Nat. Commun.* **2021**, *12*, 2354.

[22] Y. Liu, W. Hao, H. Yao, S. Li, Y. Wu, J. Zhu, L. Jiang, *Adv. Mater.* **2018**, *30*, 1705377.

[23] T. Schultz, T. Lenz, N. Kotadiya, G. Heimel, G. Glasser, R. Berger, P. W. M. Blom, P. Amsalem, D. M. de Leeuw, N. Koch, *Adv. Mater. Interfaces* **2017**, *4*, 1700324.

[24] H. Ago, T. Kugler, F. Cacialli, W. R. Salaneck, M. S. P. Shaffer, A. H. Windle, R. H. Friend, *J. Phys. Chem. B* **1999**, *103*, 8116.

[25] J. Rivnay, M. Ramuz, P. Leleux, A. Hama, M. Huerta, R. M. Owens, *Appl. Phys. Lett.* **2015**, *106*, 043301.

[26] R. Di Pietro, D. Fazzi, T. B. Kehoe, H. Sirringhaus, *J. Am. Chem. Soc.* **2012**, *134*, 14877.

[27] a) J. Huang, S. Somu, A. Busnaina, *Nanotechnology* **2012**, *23*, 335203; b) A. Ringk, X. Li, F. Gholamrezaie, E. C. P. Smits,

A. Neuhold, A. Moser, C. Van der Marel, G. H. Gelinck, R. Resel, D. M. de Leeuw, P. Strohriegl, *Adv. Funct. Mater.* **2013**, *23*, 2016.

[28] B. Analui, A. Hajimiri, *IEEE J. Solid-State Circuits* **2004**, *39*, 1263.

[29] J. Wu, X. Rui, C. Wang, W.-B. Pei, R. Lau, Q. Yan, Q. Zhang, *Adv. Energy Mater.* **2015**, *5*, 1402189.

[30] D. Kiefer, R. Kroon, A. I. Hofmann, H. Sun, X. Liu, A. Giovannitti, D. Stegerer, A. Cano, J. Hynynen, L. Yu, Y. Zhang, D. Nai, T. F. Harrelson, M. Sommer, A. J. Moulé, M. Kemerink, S. R. Marder, I. McCulloch, M. Fahlman, S. Fabiano, C. Müller, *Nat. Mater.* **2019**, *18*, 149.

[31] H. Alnoor, A. Elsukova, J. Palisaitis, I. Persson, E. N. Tseng, J. Lu, L. Hultman, P. O. Å. Persson, *Mater. Today Adv.* **2021**, *9*, 100123.